\documentclass[aps,preprint]{revtex4}
\usepackage [dvips]{graphicx}
\begin{document}

\title {\bf  NUCLEAR PAIRING: SURFACE OR BULK?}

\author{\rm
N. Sandulescu $^{a,b}$,
P. Schuck $^{b}$,
X. Vi$\tilde{n}$as $^{c}$ }

\affiliation {\rm
  $^{a)}$~ Institute of Physics and Nuclear Engineering,
           76900 Bucharest, Romania\\
  $^{b}$~  Institut de Physique Nucl\'eaire, Universit\'e Paris-Sud,
           F-91406 Orsay Cedex, France \\
  $^{c}$~  Departament d'Estructura i Constituents de la Materia,
           Universitat de Barcelona, Av. Diagonal 647, E-08028 Barcelona, Spain }

\vskip 1.0cm

\begin{abstract}

We analyse how the spatial localisation properties of pairing correlations
are changing in a major neutron shell of heavy nuclei.
It is shown that the radial distribution of the pairing density
depends strongly on whether the chemical potential is close to
a low or a high angular momentum level and  has very little
sensitivity to whether the pairing force acts in the surface
or in the bulk. The averaged pairing density over one major shell
is however rather flat, practically independent of the pairing
force. Hartree-Fock-Bogoliubov calculations for the isotopic 
chain $^{100-132}$Sn are presented for demonstration purposes.

\end{abstract}

\maketitle

\section{Introduction}

 There is an ongoing debate whether pairing in nuclei is concentrated
preferentially in the bulk or in the surface. Quite a few authors plead
for a surface enhancement of nuclear pairing \cite{tajima,fayans}.
This believe is based mostly on two arguments. First, pairing is
concentrated around the Fermi energy and in heavy nuclei levels at
the Fermi energy are dominated by high $ l$-values with corresponding
wave functions peaked at the surface.
Second, the idea of surface dominance of pairing was probably mostly
fostered from the Local Density Approximation (LDA) picture.
Indeed, from Figure 3 of Ref. \cite{garrido}, where the neutron gap at
the Fermi surface is shown in symmetric nuclear matter as a function of $k_F$,
calculated once with a realistic bare force and once with the
Gogny D1S-force, one can see that, adopting the LDA,  the gap
in the interior of a nucleus should be very small, while in the surface
it should pass through a huge peak over 2.0 MeV high. On the other
hand, using LDA one obtains on average a reasonable agreement for the
correlation energy in comparison with quantal calculations
\cite{kucharek}. However, to conclude from this fact that the LDA provides
also a good approximation for the r-dependence of $\Delta$ in nuclei is a
little premature since one knows that Thomas-Fermi theory, on which
LDA is based, yields local quantities which must be interpreted as
distributions, useful under integration but locally quite erroneous
\cite{rs}.

Actually, the two arguments presented above are related in fact to two
different aspects of pairing in finite nuclei. Thus, the first argument
refers to the radial distribution of pairing correlations, i.e.,
of Cooper pairs formed upon various single-particle states.
Commonly, the spatial localisation of pairing correlations is described
by the pairing density. In finite nuclei the pairing density is
ascribed to the radial form factor of pair transfer reactions
\cite{bm,ov,ksg}. These reactions probe the strength of pairing correlations
as manifested in pair-vibration and pair-rotation modes \cite{broglia}.
On the other hand, the pairing field $\Delta(r)$, on which the
second argument is based, provides not only informations
about the localisation of pairing correlations but also on the 
properties of the effective pairing force. This is evident from the
fact that, by definition, $\Delta(r)$ is given by the convolution
between the pairing force and the pairing density. 
As it will be shown in this paper, in finite
nuclei the radial dependence of the pairing density is rather
insensitive to the type of the pairing force. Consequently, in finite
nuclei $\Delta(r)$ and the associated global quantities (e.g.,pairing
gaps) carry in fact informations related essentially to the effective
paring force and much less on Cooper pairs localisation.

An indication that pairing correlations may not be tremendously surface
peaked came already from the study of pairing density in half infinite nuclear
matter with the Gogny force \cite{fs}. There, the peaking of the pairing density
at the surface was only very moderate. Looking at Figure 1 of Ref.\cite{fs}
it is even not evident whether the peaking is not a pure consequence of
Friedel-oscillations \cite{friedel}. For finite nuclei  only very few
calculations of the pairing density as a function of the radius
are available \cite{dobaczewski}. These calculations were focused mainly
on how the pairing density is changing
passing from one major shell to the other, up to the neutron dripline.
The scope of the present paper is to give a more systematic
investigation of the localisation properties of the
pairing density in one major shell.
Thus, it will be shown that the bulk versus surface localisation of
pairing correlations can change very much  already in one major shell,
depending on whether the chemical potential is close to a low or a high
angular momentum level. This behaviour is illustrated here for the chain
of $Sn$-isotopes with the neutron number between N=50 and N=82. The study
is performed in the Hartree-Fock-Bogoliubov approach and using
surface versus bulk dominated pairing interactions. The general framework of the calculations is
described in Section II. Then, in Section III, we discuss the radial
distribution of pairing correlations as provided by the HFB calculations.
Finally in Section IV we present our conclusions.

\section{Formalism: Hartree-Fock-Bogoliubov Approach}

 In superfluid Fermi liquids, pairing correlations are usually
characterized
 by the  "condensate" wave function \cite{noziere}. For the case of $^1S_0$
 pairing the condensate wave function, referred below as to the pairing
density,
 is given by:
\begin{equation}
\kappa({\bf r})= \langle \psi({\bf r},s=1/2) \psi({\bf r},s=-1/2) \rangle ,
\end{equation}
where the operator $\psi({\bf r},s=1/2)$ annihilates a nucleon
in the point ${\bf r}$ and having the spin projection $s=1/2$.
The local pairing density defined above describes the center of mass
distribution of the Cooper pairs inside the superfluid.
For translationally invariant superfluid systems at zero temperature,
$\kappa({\bf r})$ is a constant field with the modulus equal to $\sqrt n_0$,
where $n_0$ is the number of the condensed (Cooper) pairs \cite{noziere}.
In large superfluid systems, in which the condensate is macroscopically
occupied, $\kappa({\bf r})$ can still be regarded as a classical coherent
field with negliglibe variations over distances greater than the coherence length.
However, when the range of  pairing correlations (associated to the mean square
radius of Cooper pairs) is comparable with the size of the system, as happens
in finite nuclei, the local variations of the pairing density are important
and should  be calculated from a quantum equation of motion of the
Bogoliubov-de-Gennes type \cite{gennes}. In
finite nuclei for such a task one usually employs the
Hartree-Fock-Bogoliubov equations \cite{rs}, in which both
the mean field and the pairing field are calculated self-consistently.

 In this study the HFB equations are solved by imposing spherical symmetry
 and employing zero-range forces in both the particle-hole and
 particle-particle channels. Under these conditions the radial HFB
 equations have the following form:
\begin{equation}
\begin{array}{c}
\left( \begin{array}{cc}
h(r) - \lambda & \Delta (r) \\
\Delta (r) & -h(r) + \lambda
\end{array} \right)
\left( \begin{array}{c} \mathfrak{U}_i (r) \\
 \mathfrak{V}_i (r) \end{array} \right) = E_i
\left( \begin{array}{c} \mathfrak{U}_i (r) \\
 \mathfrak{V}_i (r) \end{array} \right) ~,
\end{array}
\label{1}
\end{equation}
\\
where $h(r)$ and $\Delta(r)$ are, respectively, the mean field hamiltonian
and pairing field, while $\lambda$ is the chemical potential. All basic
quantities such as particle density $\rho(r)$ and pairing density
$\kappa(r)$ are expressed in terms of the upper and lower components of
the HFB wave function. Thus:
\begin{equation}
\rho(r) =\frac{1}{4\pi} \sum_{i} (2j_i+1) \mathfrak{V}_i^* (r)
\mathfrak{V}_i (r) ,
\label{2}
\end{equation}
\begin{equation}
\kappa(r) = \frac{1}{4\pi} \sum_{i} (2j_i+1) \mathfrak{U}_i^* (r)
\mathfrak{V}_i (r) .
\label{3}
\end{equation}

In the present HFB calculations we use in the particle-hole channel
a Skyrme-type force,i.e., SLy4 \cite{sly4}. Hence the mean field has
the standard expression in terms of single-particle densities
\cite{vautherin,dft}. For the pairing interaction we use two forces,
i.e., a pure $delta$-force and a density- dependent $delta$ (DDD) -force.
For the latter one we take the form \cite{bertsch}:
\begin{equation}
V (\mathbf{r}-\mathbf{r^\prime}) = V_0 [1 -
\eta(\frac{\rho}{\rho_0})^{\alpha}]
\delta(\mathbf{r}-\mathbf{r^\prime})
\equiv V_{eff}(\rho(r)) \delta(\mathbf{r}-\mathbf{r^\prime}).
\end{equation}
The pairing interaction acts upon the pairing density through the
pairing field, which for the DDD-force is given by:
\begin{equation}
\Delta(r) = V_{eff}(\rho(r)) \kappa (r).
\end{equation}

Due to the divergencies associated to a zero range pairing force, the HFB
calculations should be performed with an energy cut-off or by using more
sophisticated regularisation procedures \cite{bulgac}.
In the present calculations the energy cut-off and the strength of the
DDD- force are related to each other through the
scattering length of the di-neutron system\cite{bertsch}.
The density dependence of the DDD- force is fitted so as to reproduce
approximatively the pairing gap in nuclear matter provided by the
Gogny D1S- force \cite{garrido}.
Guided by these prescriptions we have taken for the DDD- force
the following parameters: $V_0$= -430 MeV fm$^{-3}$, $\eta$=0.45
and $\alpha$=0.47. For the cut-off energy (in the quasiparticle
spectrum) we use a value equal to 60 MeV. In the HFB calculations
with the $delta$-force we have taken a strength $V_0$= -220 MeV fm$^{-3}$
and we have used the same cut-off energy.
With this value of the strength we get for Sn isotopes approximatively
the same pairing energies as for the DDD -force.

\section{Results: Pairing Localisation in Sn Isotopes}

Before we start analysing the local properties of pairing correlations, we
first present shortly the global quantities  which charaterize the
HFB solution. Thus, in Table I is shown how the chemical potential
$\lambda$, the averaged gaps and the occupation probabilities of
single-particle states are evolving by filling the major shell N=50-82.
The changes of particle density with the neutron number are shown in Figure 1a.
These changes can be easily traced back to the occupation probabilities of
single-particle wave functions shown in Figure 1b.  Thus, one notices the
progressive increase of the particle density at small distances, produced
by the filling of the state $3s_{1/2}$.

\begin{table}
\begin{center}
\begin{tabular}{|c|c|c|c|c|c|c|}
\hline
A           &  104  &  108   &  114    &  116    &   124   &   128  \\
\cline{1-7}
$\lambda$   & -10.9 &  -10.0 &  -8.97  &  -8.64  &  -7.3   &  -6.7  \\
\cline{1-7}
$\langle \Delta \rangle $
            &  1.4  &  1.77  &  1.91   &  1.88   &  1.6    &  1.3    \\
\cline{1-7}
$2d_{5/2}$ (-10.1)  &  0.63 &  0.73  &  0.88   &  0.91   &  0.97   &  0.98   \\
\cline{1-7}
$1g_{7/2}$ (-9.7) &  0.17 &  0.28  &  0.65   &  0.74   &  0.94   &  0.97   \\
\cline{1-7}
$3s_{1/2}$ (-8.8) & 0.097 &  0.16  &  0.43   &  0.54   &  0.89   &  0.95   \\
\cline{1-7}
$2d_{3/2}$ (-8.3) & 0.07  &  0.11  &  0.30   &  0.4    &  0.83   &  0.93   \\
\cline{1-7}
$1h_{11/2}$ (-6.7) & 0.03  &  0.05  &  0.11   &  0.14   &  0.44   &  0.7    \\
\hline
\end{tabular}
\end{center}
\caption{Results of HFB calculations for $Sn$ isotopes obtained with
the DDD-force. $A$ is the atomic mass, $\lambda$ is the chemical potential
(in MeV) and $\langle \Delta \rangle $ is the averaged pairing gap (in
MeV). The latter is calculated by convoluting the pairing field with
the pairing density. In the next 5 rows are given the occupations
probabilities for the single-particle states of the valence shell.
For each of these states is shown in the bracket the Hartree-Fock
energy corresponding to the midshell isotope $^{116}$Sn.}
\label{table1}
\end{table}

\begin{figure}[h]
\begin{center}
\includegraphics*[scale=0.35,angle=-90.]{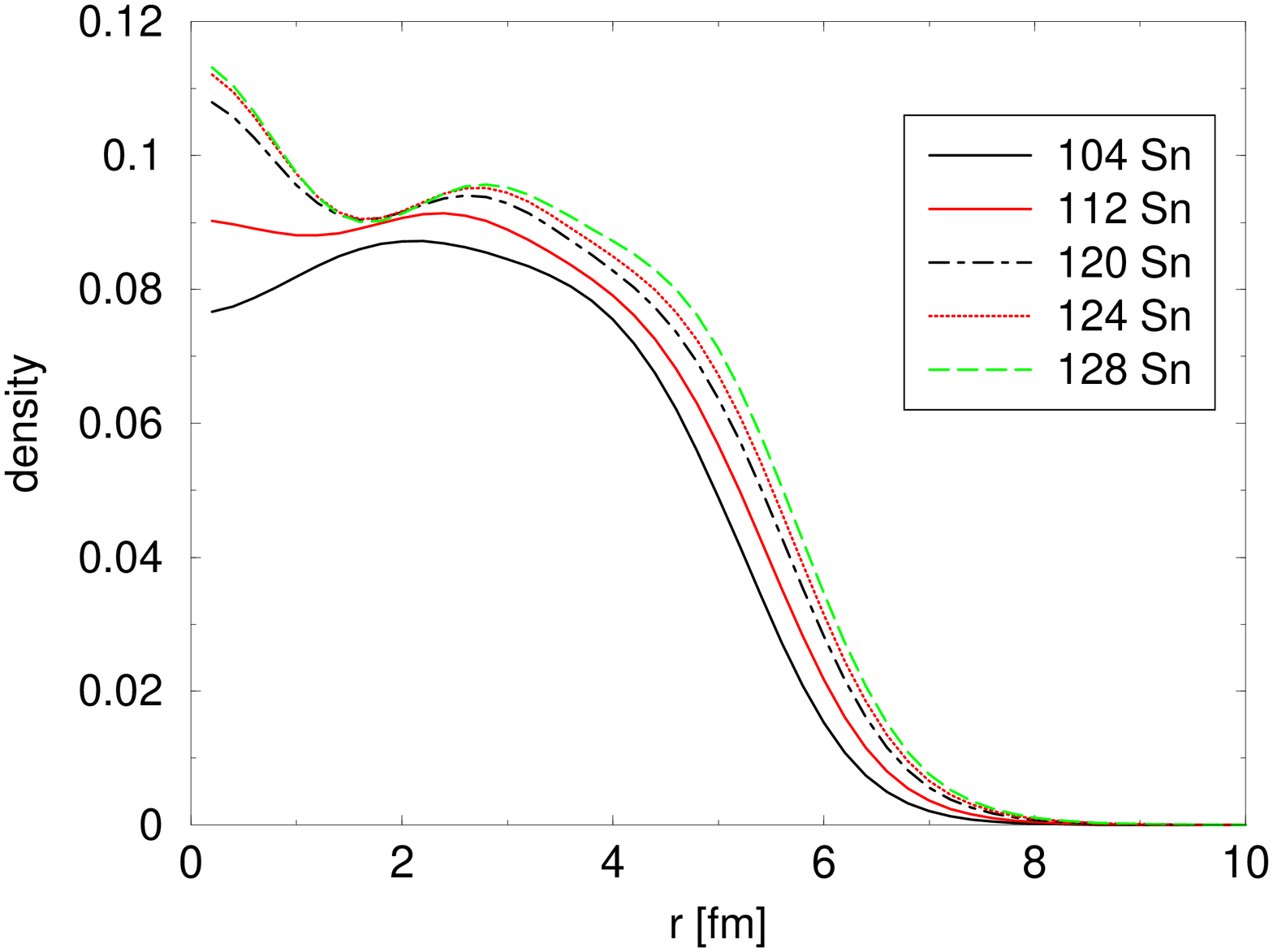}
\includegraphics*[scale=0.35,angle=-90.]{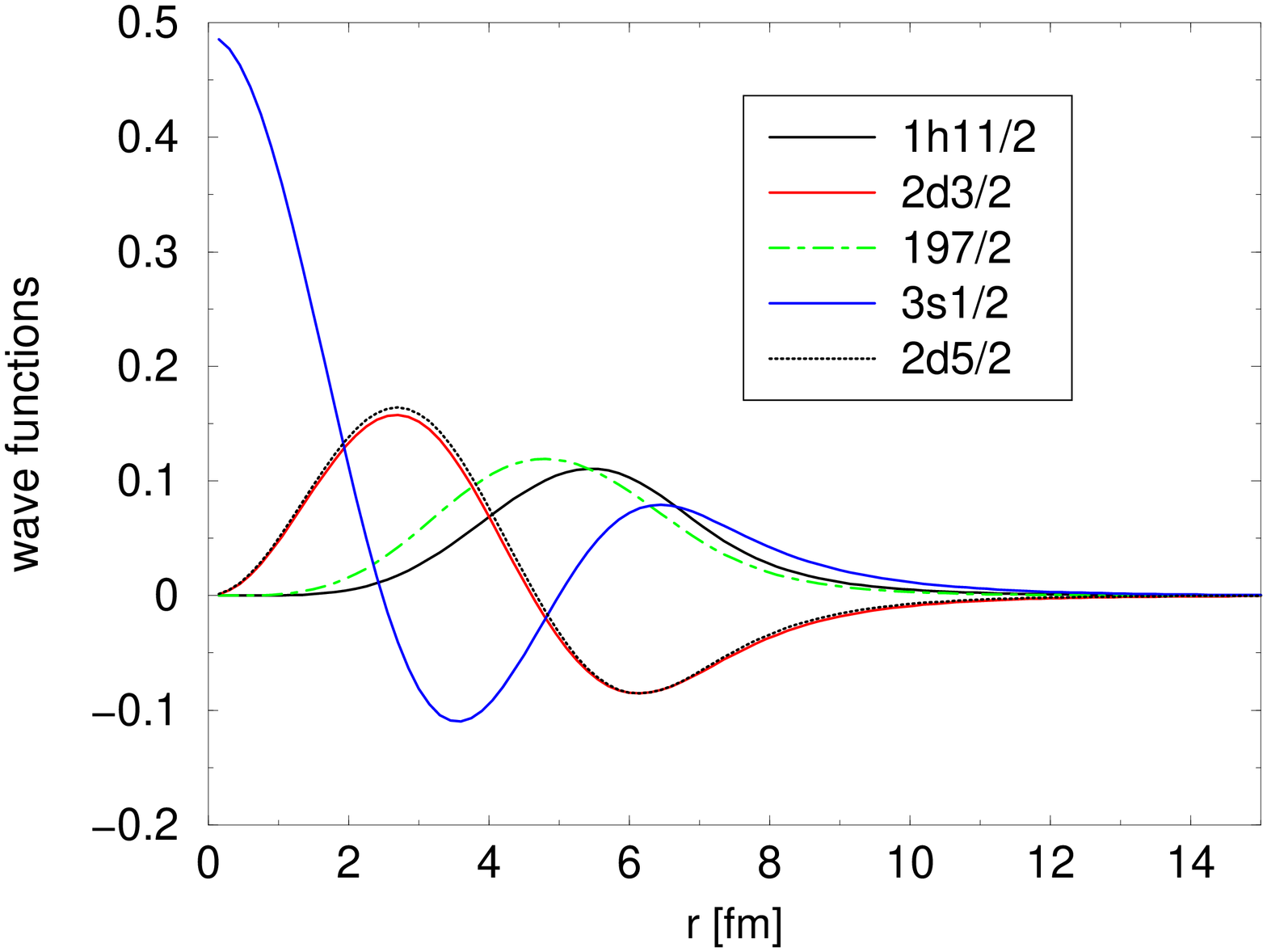}
\caption{ The particle density (a) for various Sn isotopes and the
Hartree-Fock wave functions (b) corresponding to the valence
shell of $^{116}$Sn.}
\end{center}
\end{figure}

Next we discuss how the localisation properties of the pairing density
are changing with the filling of the major shell N=50-82.
The results for $\kappa(r)$ obtained by using the two zero- range
pairing forces introduced in the previous section are shown in Figure 2.

\begin{figure}[h]
\begin{center}
\includegraphics*[scale=0.35,angle=-90]{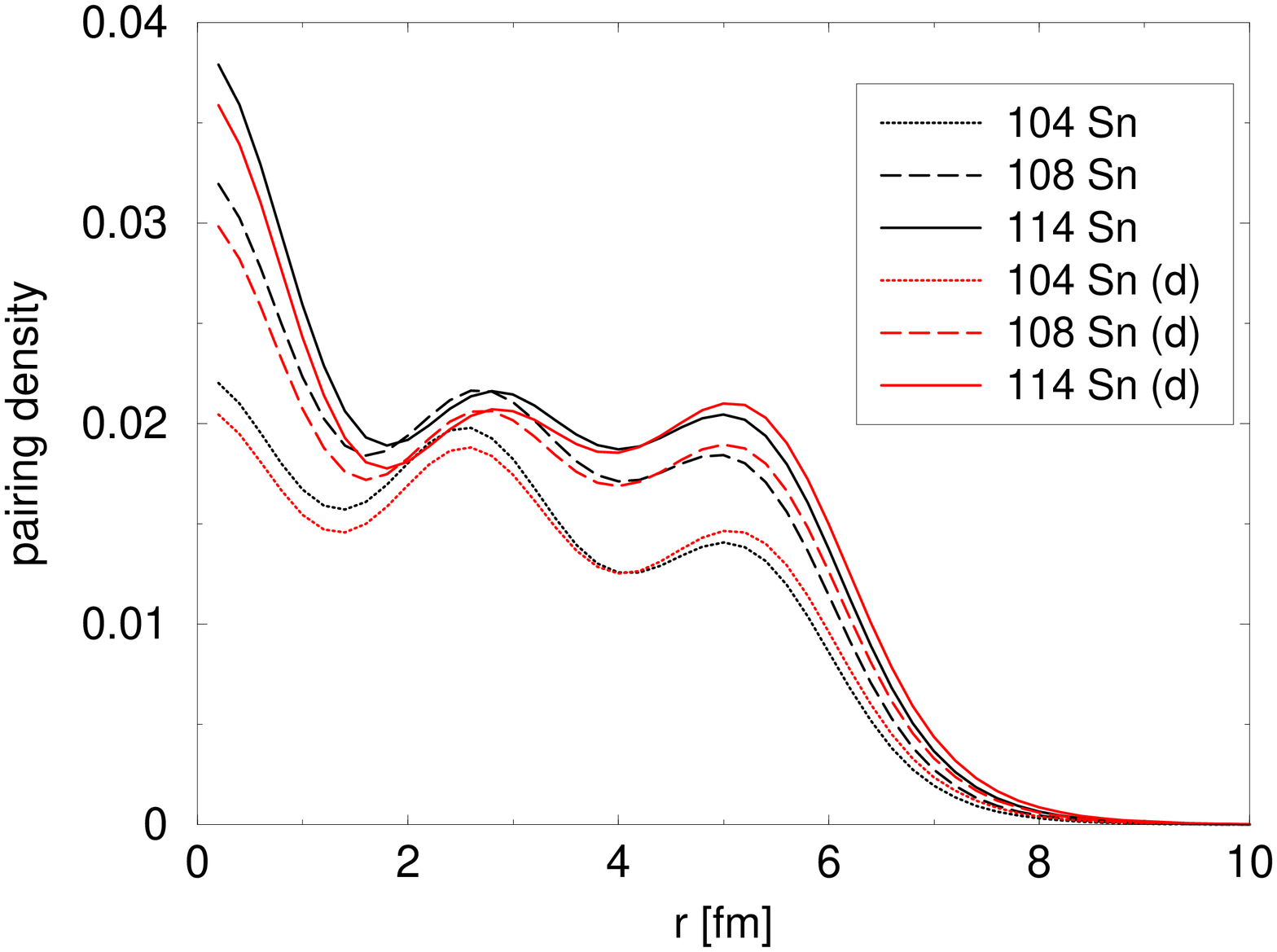}
\includegraphics*[scale=0.35,angle=-90.]{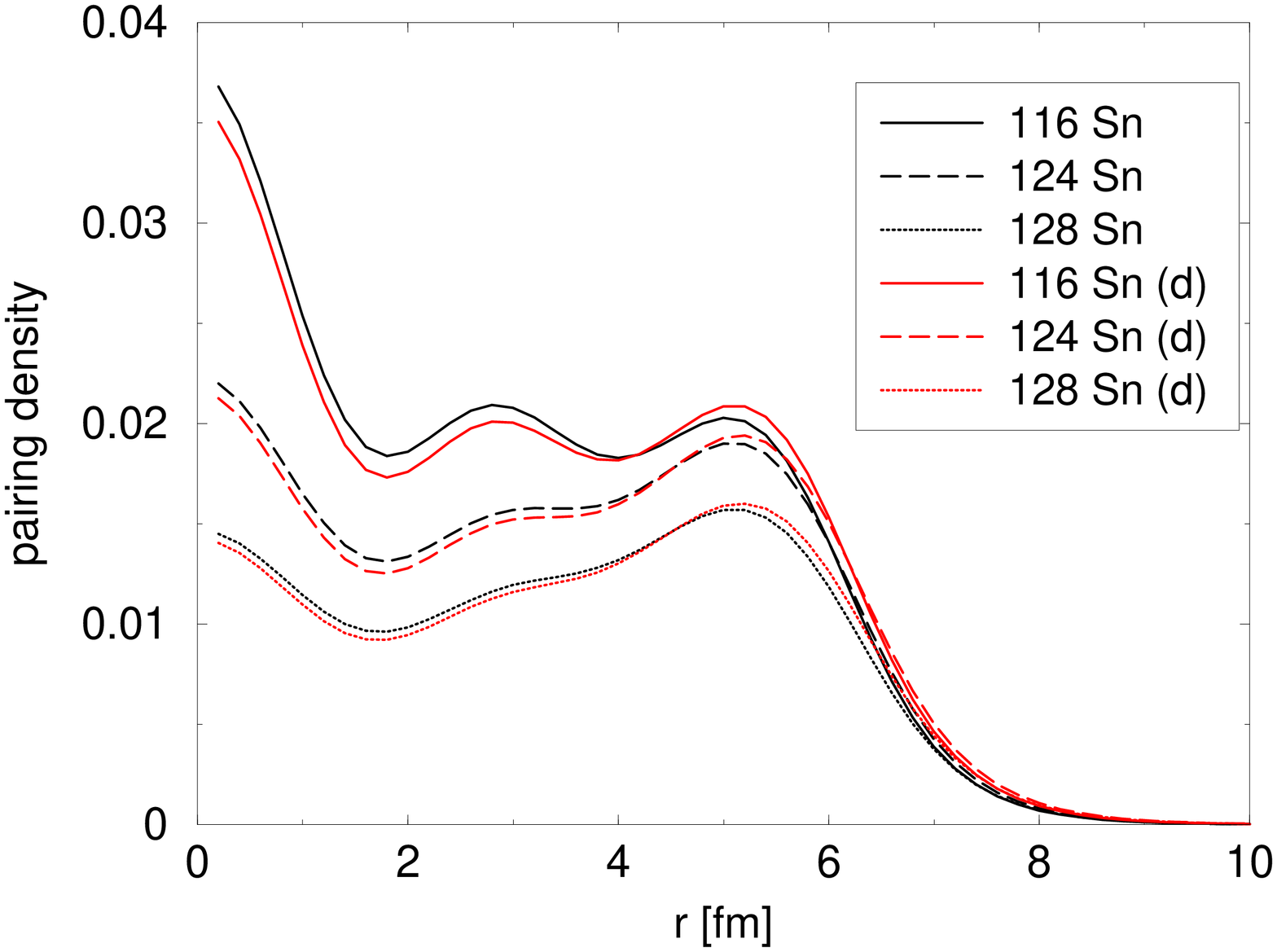}
\end{center}
\caption{ Neutron pairing densities $\kappa(r)$ (in fm$^{-3}$) for
             Sn isotopes calculated in the HFB approach.
             The black (read) curves correspond to the $delta$-force
             (density-dependent $delta$-force).}
\end{figure}

First, we notice that for both pairing forces  $\kappa(r)$ is changing
quite strongly passing from A = 104 to A = 128. Thus, we see that at the
beginning of the shell, when  the contribution of the states $2d_{5/2}$
and $1g_{7/2}$ is dominant, the pairing density is slightly larger
inside the nucleus than in the surface. Then, at midshell, the pairing
density becomes in average almost flat for an extended region from
about 1 fm to 5 fm. We also notice a large bump close to the center
of the
nucleus produced by the neutrons distributed in the state $3s_{1/2}$.

In the second half of the major shell the chemical potential
becomes closer to the intruder state $1h_{11/2}$, which starts
progresively to dominate the structure of $\kappa(r)$. Since
the wave function of $1h_{11/2}$ is localized in the surface
region (see Figure 1b), the pairing density is getting also a bumb
around 5 fm. This bumb becomes more pronounced towards the end
of the shell. On the other hand, one can see that towards the
end of the shell the pairing density decreases rather strongly
in the center of the nucleus. This behaviour reflects the
smaller contribution of the state $3s_{1/2}$ towards N=82.

A somewhat surprising but important finding, shown in Figure 2,
is that the radial structure of $\kappa(r)$ is changing very
little with the density dependence of the $delta$-force .
Through this dependence
the strength of the pairing force is reduced inside the nucleus
compared to the surface region. Consequently, the relative
contribution of the state $3s_{1/2}$ to pairing correlations
is suppressed. However, because of its small degeneracy compared to
the other states, this suppression has not important consequences
on $\kappa(r)$, as can be seen in Figure 2.

\begin{figure}[h]
\begin{center}
\includegraphics*[scale=0.40,angle=-90.]{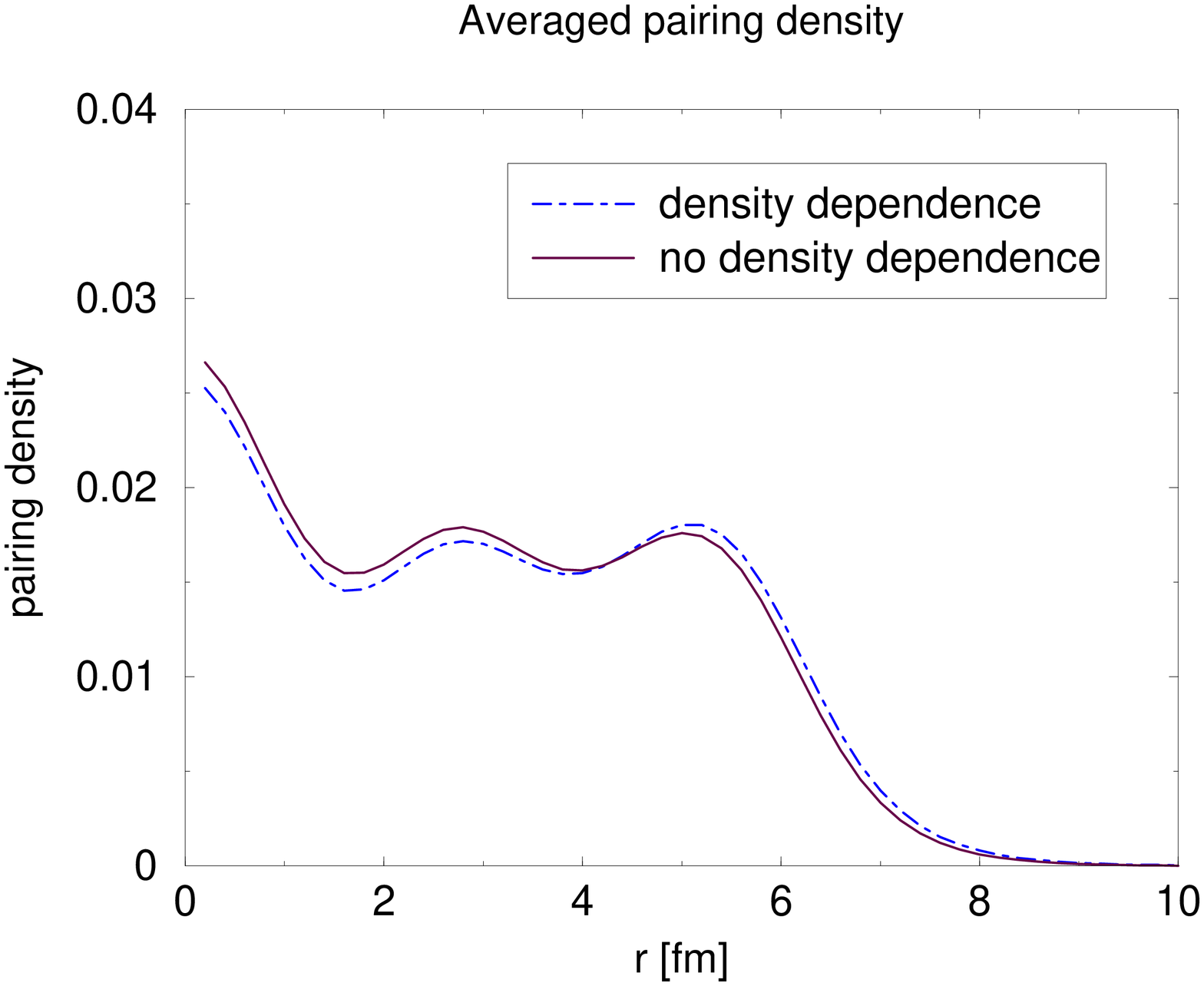}
\caption{ Averaged pairing density (in fm$^{-3}$) calculated in the
          HFB approach. The solid (dot-dashed) curve corresponds to
          $delta$-force (density-dependent $delta$-force).}
\end{center}
\end{figure}

As we have seen above, the radial dependence of the pairing density in
Sn-isotopes is dominated by the individual shells structure.
In order to reveal a generic behaviour one should follow the old
idea of Strutinsky \cite{strutinsky} and average the pairing
density over one major shell. Here we have taken the crude arithmetic
average over all even Sn isotopes, leaving the study of more refined
averaging methods for future work. The result is shown in Figure 3.
We remark that on average the pairing density $\overline{\kappa}(r)$ is
quite flat over the bulk.
Since we are considering an even parity shell, we have at the origine a
quite pronounced bump from the s-wave. Had we considered an odd parity
shell, we certainly would see a corresponding hole at the origin
\cite{hole}. In Figure 3 we also see  that in spite of the fact
that the density dependent force yields, with respect to the pure
$\delta$-force, slight surface enhancement and volume depression,
the difference induced upon $\overline{\kappa}(r)$ by the two
pairing forces is practically insignifiant. 

 In order to further understand what happens, let us take the midshell
situation and approximate the $uv$ factors in the BCS expression of
kappa with a unique constant equal to one, i.e.,  
$\kappa(r)=\frac{1}{4\pi}\sum_iu_iv_i|\phi_i(r)|^2
 \approx \frac{1}{4\pi} {\sum_i}'|\phi_i(r)|^2$, 
where the prime indicates that the sum
 runs over the major shell only and $\phi_i(r)$ are the single-particle
 wave functions plotted in Figure 1b. The result of this approximation is
 shown in Figure 4. By summing in $\kappa$ all single-particle states with
 the  same $uv$ factor one overestimates the contribution of those
 single-particle states which are far from the chemical potential,
 mainly of the states $2d_{5/2}$ and $h_{11/2}$. This fact produces
 the extended tail seen in Figure 4. Apart from that, we can see that
 this rough approximation of $\kappa$  follows rather well the flat
 radial  structure of the HFB curve.

\begin{figure}[h]
\begin{center}
\includegraphics*[scale=0.45,angle=-90.]{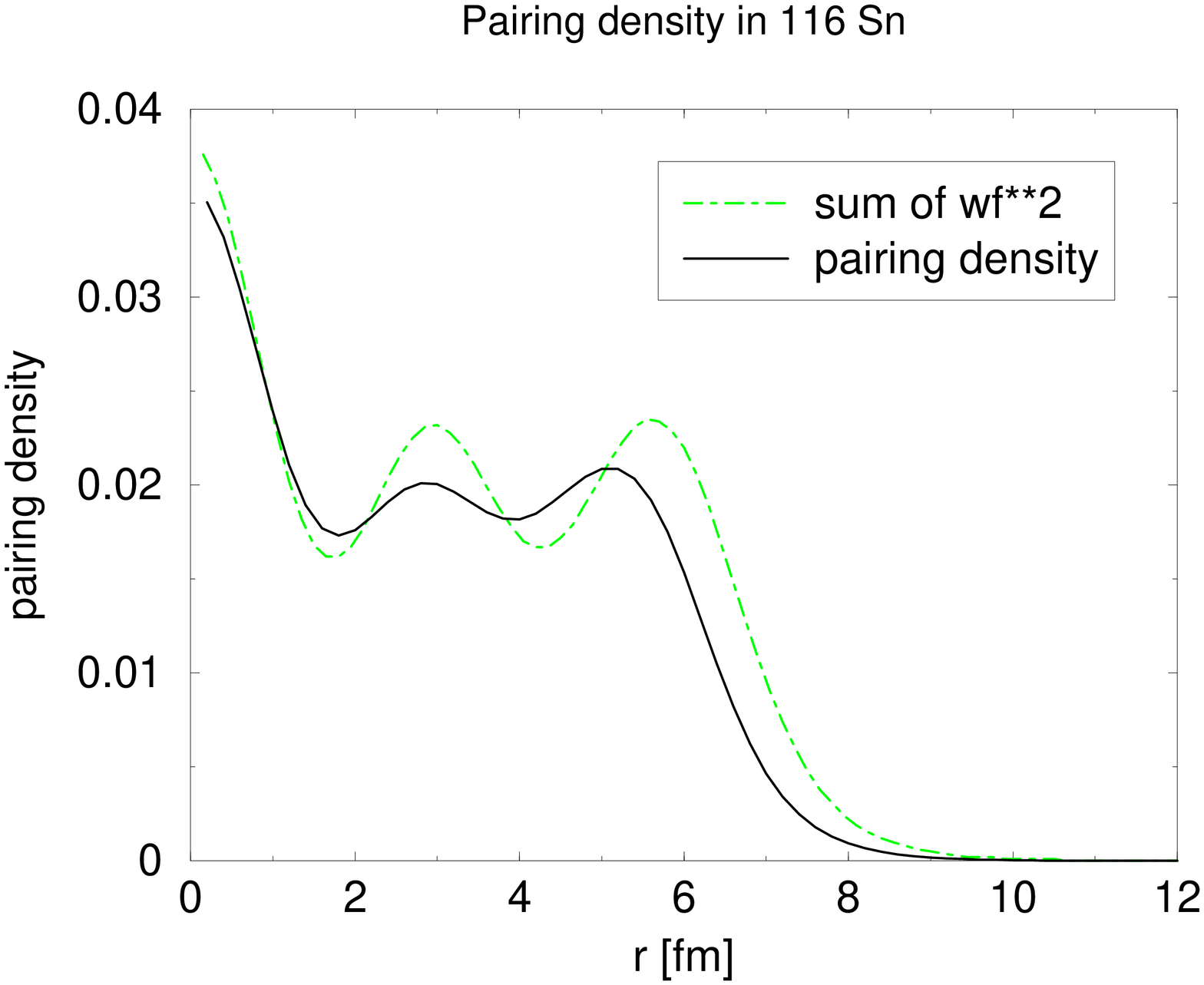}
\caption{ Pairing density in $^{116}$Sn. The solid line is the HFB
          result. The dotted line correspond to the summation of
          the square of single-particle wave functions (see text).}

\end{center}
\end{figure}

From all these results it appears that in a particular nucleus the radial
distribution of the pairing correlations, given by the pairing density,
is determined essentially by the localisation of single-particle states
which are closer to the chemical potential and much less by the type
of the pairing force. The dependence of pairing density profile on the
filling of the major shells and its rather small sensitivity to the
assumption made on the pairing force can be eventually checked
in pair transfer reactions.

The quantity which, by definition, depends explicitly on the
assumption made on the type of pairing force is the pairing field.
This dependence is clearly seen in Figure 5, where the averaged
pairing fields corresponding to the $delta$-force and the DDD-force
are plotted. Hence, indications about which type of pairing force might
be more appropiate in finite nuclei can eventually be extracted
from
quantities related to the pairing field, e.g., the pairing gaps.
Thus, according to Ref.\cite{dns} a DDD-force with
$\eta=1/2$ and $\alpha=1$ would give better results for the odd-even
mass staggering than a $\delta$-force or a DD-force with 
$\eta=1$ and $\alpha=1$. However, a pure $\delta$- force apparently
gives for the one- and the two- neutron separation energies almost
the same results as a DDD-force \cite{yb}. Therefore, from these
calculations one cannot draw yet a definite conclusions on how much
stronger
the pairing force should be on the surface compared to the bulk.
On the other hand, the fact that the pairing force should have some
surface peaking is suggested by all bare interactions, which
are more attractive for small momenta than for large ones
(see also the $V_{low-k}$ force based on realistic bare
nucleon-nucleon interactions \cite{schwenk}).

\begin{figure}[h]
\begin{center}
\includegraphics*[scale=0.50,angle=-90.]{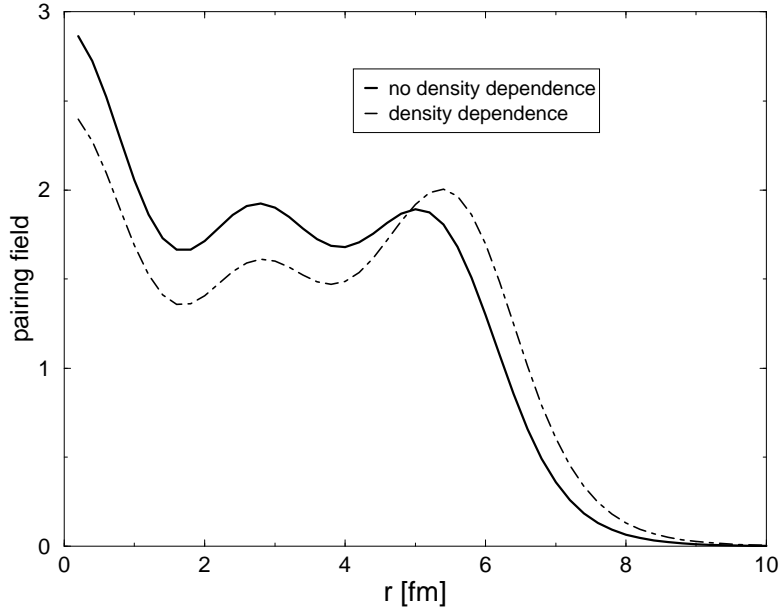}
\caption{ Averaged pairing field (in MeV) calculated in the
          HFB approach. The solid (dot-dashed) curve correspond to
          $delta$-force (density-dependent $delta$-force).}
\end{center}
\end{figure}

\section{Summary and Conclusions}

In this work we have investigated the radial distribution of
pairing correlations in the major neutron shell N=50-82
of Sn isotopes. The localisation of pairing correlations, given
by the  pairing density, is calculated in the framework of HFB
approach and using two effective pairing interactions, i.e.,
a pure $\delta$-force and a $\delta$-force with a density dependent factor
which enhances the strength of the interaction at low density,
i.e., in the surface region. It was found that the pairing density
changes strongly going from one end to the other of
the major shell N=50-82, depending on whether the chemical potential
is close to a level with low or high l-value. However, the
differences with respect to the two pairing forces stayed
insignificant.

Then, in order to obtain some generic behaviour
of the pairing density $\kappa$ in nuclei, we performed an
arithmetic average of
$\kappa$'s over one major shell. The resulting averaged pairing density,
$\overline{\kappa}$, is, apart from some oscillations, practically constant
over the whole volume besides a quite pronounced peak at the origin,
produced by the $3s_{1/2}$-state contained in the open major shell of
Sn-isotopes. The
results for $\overline{\kappa}$
show again very little sensivity to the type of the pairing force
used in the calculations.

We have also calculated the corresponding pairing fields $\Delta(r)$.
By definition, the pairing field depends explicitly on the pairing
force. Thus, even if for the DDD-force the pairing density is almost
constant over the nuclear volume, the corresponding pairing field
would get a similar surface enhancement as the DDD-force. Therefore
in finite nuclei quantities relating to the radial distribution
of the pairing field, e.g. odd-even mass differences, probes the
density dependence of the pairing forces rather than the localisation
of pairing correlations, i.e., of Cooper pairs.

Summarizing, the radial distribution of pairing correlations in finite
nuclei depends strongly on the localisation of single-particle states
which are closer to the chemical potential and much less on the surface
or bulk enhancement of the pairing force. 

The conclusions of this study are based on zero range pairing forces.
However, the results with a finite-range force of Gogny type,
which will be published in a future paper together with a
semiclassical study of nuclear pairing localisation \cite{ssvv},
show the same trends.

\end{document}